\documentstyle[epsfig,12pt]{article}
\def\thefootnote{\fnsymbol{footnote}}
\begin{document}
\begin{titlepage}
\today          \hfill 
\begin{center}
\hfill    LBNL-46903 \\
\hfill    UCB-PTH-00/34 \\
\hfill hep-th/0009214 \\

\vskip .5in
\renewcommand{\thefootnote}{\fnsymbol{footnote}}
{\Large \bf Tachyon instability and Kondo type models} \footnote{This work was supported by the Director, Office of Energy 
Research, Office of High Energy and Nuclear Physics, Division of High 
Energy Physics of the U.S. Department of Energy under Contract 
DE-AC03-76SF00098 and in part by the National Science Foundation grant PHY-95-14797.}
\vskip .50in

\vskip .5in
 Korkut Bardakci\footnote{email address: kbardakci@lbl.gov} and Anatoly Konechny\footnote{email address: konechny@thsrv.lbl.gov}

\vskip 0.5cm
{\em Department of Physics\\
University of California at Berkeley\\
   and\\
 Theoretical Physics Group\\
    Lawrence Berkeley National Laboratory\\
      University of California\\
    Berkeley, California 94720}
\end{center}

\vskip .5in

\begin{abstract}
In this paper we investigate the tachyon instability of open 
bosonic string theory applying methods of boundary conformal 
field theory. We consider compactifications  on maximal tori 
of various simple Lie algebras with a specific background 
coupled to  the string boundaries. The resulting 
world-sheet CFT is a free theory perturbed by a boundary term that is marginal 
but not truly marginal. Assuming that the theory flows to 
a nontrivial infrared fixed point that is similar to  the one in 
Kondo model, we calculate the new spectrum and some of the Green's 
functions. We find that in some of the sectors the tachyon mass 
gets lifted that can be interpreted as a result of  switching on 
appropriate Wilson lines.    Various compactifications and 
 patterns of flows are investigated.  
 
\end{abstract}
\end{titlepage}

\newpage
\renewcommand{\thepage}{\arabic{page}}
\setcounter{page}{1}
\section{ Introduction}

The perturbative vacuum in the pure bosonic string models is known to be 
unstable because of the existence of the tachyonic mode, leaving open the
possibility of  the existence of 
 another stable vacuum. This question was first investigated
for open strings
in references \cite{Bardakci1}, \cite{BardHalp}, \cite{Bardakci2}, and later, using string field theory 
techniques, in \cite{Sam1}, \cite{Sam2}, and the
indications are that there is such a vacuum.

More recently, Sen \cite{Sen} has revived
interest in this subject by his investigations of the brane-antibrane complex in
the superstring, which also has a tachyonic instability, due to the breaking of 
supersymmetry. Since then, there has been numerous papers on this subject 
(\cite{Sen2}, \cite{SenZw}, \cite{BerkSenZw}, \cite{MTaylor} and references therein). 
All of these papers reach the conclusion that in the end, tachyon condensation takes place,
and as a result, the supersymmetric vacuum is restored.

The tachyon instability can be pictured in analogy with Higgs phenomenon as resulting 
from the perturbative vacuum  being put  on top of the tachyon potential well. 
Most of the efforts in the literature were directed towards an explicit  calculation 
of the space-time tachyon potential in various approximation schemes. 
Recently   in papers \cite{Shat}, \cite{KMM} an effective tachyon Lagrangian in the first derivative 
approximation was calculated exactly by applying methods of background independent  open string field theory.

Perhaps less explicitly   but technically easier  the tachyon condensation in open string theory  can be  studied by 
methods of  boundary conformal field theory (BCFT). The general  idea is as follows. 
In the sigma model formulation of open string theory  the starting point is  
a world-sheet action of the form $S = S_{0} + S'$ where $S_{0}$ is a fixed bulk action 
specifying the closed string background that can be taken to be  the standard flat background 
and $S'$ is a boundary perturbation. For example in the open bosonic theory one can consider  
\begin{equation} \label{tach-pot}
S' = \int_{\partial \Sigma} d\tau \, V[X(\tau )]
\end{equation}
where $V[X]$ is the tachyonic field profile. 
The classical open string equations of motion follow from the requirement of the world-sheet conformal 
invariance that is ensured by vanishing of the beta functions of the world-sheet theory. 
The world-sheet RG flow to the IR induced by the boundary perturbation $S'$ thus describes approaching 
a classical solution in space-time. This classical solution in general corresponds to some intermediate 
metastable state that can be perturbed further until the system reaches the true vacuum. 
The perturbation of the form (\ref{tach-pot}) was studied in \cite{relevance}. It was shown in that paper that for 
 a particular choice of  the potential $V[X]$ (previously considered in     
\cite{Polchinski}, \cite{CalMald}, \cite{FSW}) the arising RG flow describes disintegration of D25 brane into a 
system of lower-dimensional branes. Other types of brane descend relations in  bosonic string theory 
were studied in \cite{Sen1}. It is believed that at the final point of tachyon condensation the 
theory contains no open string excitations at all. 
Thus, quite generally tachyon condensation is the process describing  a reduction of the number of open strings degrees of freedom. 
A natural measure of the number of  degrees of freedom in BCFT  is the boundary entropy 
\cite{AfflLud2}. It was shown in \cite{AfflLud3} that at least in the lowest order in conformal perturbation theory 
this quantity always decreases.

In this paper we  consider boundary  backgrounds of a more general type than (\ref{tach-pot}) that 
include nontrivial Chan-Paton factors.
 Inclusion of Chan-Paton factors  results in   a  path integral measure on the world-sheet with boundary  that is weighted by 
$$
e^{-S_{0}}{\rm Tr \, Pexp} \Bigl( -\int_{\partial \Sigma} d\tau \, V \Bigl)
$$
where $V$ in general depends on $X$ and its derivatives  and carries  matrix indices corresponding to Chan-Paton 
degrees of freedom.

The     boundary probes we consider are constructed as follows.  
We first compactify a number
of extra dimensions at self-dual radii, thereby generating an internal
affine algebra. This algebra is then coupled to external Chan-Paton factors
in various representations of subgroups of the affine algebra.  
In this fashion, the direct product of the external and internal groups is broken
down to the diagonal subgroup, and the resulting models are various generalizations of
 the Kondo model. Although the 
boundary interaction is mediated by an operator whose conformal dimension is marginal, 
the models  are not conformal for arbitrary coupling constant.
Following the work of Affleck and collaborators \cite{Affleck}, \cite{AfflLud1}, 
\cite{AfflLud2}, \cite{AfflLud3} on the Kondo model, we
assume that under the renormalization group flow, the coupling constant flows
into a unique value demanded by conformal invariance. Consequently, one obtains
a conformal model with a specific boundary interaction, which
  can be solved exactly by various methods. The resulting physics is well known from the
example of the Kondo model: The boundary spin (Chan-Paton factors) fuses with
the internal spin and leads to a reordering of the levels. A possible relation of 
a different sort between   Kondo-like problems  and the tachyon condensation 
on  branes was suggested in \cite{GNS}.

The method of solution we choose makes use of what we call the singlet operators,
which are combinations of the Chan-Paton wavefunction with the standard
 twist operators (e.g. see \cite{Ginsparg}),
 which explicitly
exhibit the fusion of the boundary ``spin'' into the bulk. These are most 
conveniently constructed in terms of the bosonic string operators. In many cases,
simple fermionic representations also exist, and they are also of some interest.
 In this way, we are able
to treat a variety of problems, involving both single and double boundaries, and
several different compactifications, resulting in symmetry groups ranging all the
way from $SU(2)$ to $SO(32)$. We consider several examples, and for these, we compute
the spectra, and exhibit both the states and the vertex operators for low lying levels.
In the case of $SU(2)$, we also work out some of the 4 point amplitudes.
The result of these calculations can be summarized very simply: The momenta in the
compactified dimensions, which were integerly spaced in natural units at the beginning,
flow into half integer values. In the fermionic language when it is applicable, such as
in the case of $SO(4)$,
this often means a change in the fermionic boundary conditions: The NS fermions flow
into the R fermions. All of this is in agreement with the results of Affleck and
collaborators \cite{Affleck} - \cite{AfflLud3}. 
The mass of the original tachyon is lifted up, in some cases all
the way up to zero, and in other cases only part of the way. Unfortunately, consistency
requires that integer valued momenta must always be present, and so, although the 
tachyon mass may be lifted in some channels, it will reappear unchanged in other
channels. 

Next, we generalize from $SU(2)$ to higher groups such as $SO(4)$, $SO(8)$, $SO(16)$ and $SO(32)$
by compactifying in increasing number of dimensions. In these cases, one has more
options with the choice of the group representations for the Chan-Paton wave functions,
 and compared to $SU(2)$, more
complicated patterns of shift in the spectrum emerge. In the fermionic picture, when it
exists, the general rule is a flow from NS to Ramond fermions.
 We also consider the $SU(n)$ groups,
and discuss the case $SU(3)$ in detail, pointing out the simple description of what is
happening in the T-dual picture in terms of D-branes. 
We give a further discussion of   a possible brane interpretation of the 
RG flows we considered in  the last section of the paper.  
We end  with some speculations
about flowing into a superstring. The most favorable case is the full compactification
of 16 coordinates, resulting in the $SO(32)$ group. In the fermionic picture, after the flow
from NS to Ramond fermions, one has the full world sheet fermionic structure needed to
construct the superstring.
 This possibility of the bosonic string leading to
a superstring has been suggested before \cite{BerkVafa}, \cite{BergGab} (see \cite{Matsuo} 
for some of the more recent suggestions) and remains to see whether it can be fully realized.

The paper is organized as follows: In section 2, we consider the compactification of one extra
dimension of  the open bosonic string at self dual radius, with the resulting $SU(2)$ current algebra.
 In addition, we  introduce external $U(2)$ Chan-Paton factors, and couple the two SU(2)'s into
the diagonal subgroup.
  We then study in detail the
the conformal theory into which this model flows.
 In section 3, we construct the corresponding
vertex operators and compute some four point amplitudes explicitly, and we also show how to
extend the computation to higher amplitudes.  These calculations show that
the originally integerly moded momenta of the compactified dimension flow into
half integer values. In section 4, we discuss the compactification of two extra dimensions, with
the resulting $SU(2)\times SU(2)$ or $SO(4)$ current algebra. Coupling one of the $SU(2)$'s
to Chan-Paton factors, we again find a shift of half a unit in the bosonic momenta. We also 
reconsider the same problem using fermionic fields, and show that the half integer shift in the
bosonic momenta corresponds to a flow from the NS fermions to Ramond fermions in the fermionic
picture. These results agree with the well-known results in the Kondo model. In section 5, we
consider the compactification of extra dimensions all the way up to 16, and the resulting
more complicated conformal models based on bigger groups are studied by the same methods as
before.  Finally, in section 6, we list some interesting problems that our work naturally
suggests.

\section{Compactification of one  dimension} 

In string theory there are two different ways of introducing gauge 
symmetry. One way is via Chan-Paton factors  putting  charges, called quarks, on the ends of the open string.
The other way is using current algebras when the charge is distributed along the string. 
At the source of the second possibility is the effect of gauge symmetry enhancement emerging at special  
compactification radii.  
The simplest  example  of this kind of  compactification is  the bosonic  string compactified   
on a circle of the self-dual radius $R=\sqrt{\alpha'}$. 
In this case the massless spectrum  contains additional states forming  $SU(2)_{L}\times SU(2)_{R}$ 
 multiplets. More precisely in the closed string case we have gauge bosons living in the uncompactified 25 
dimensions and a $({\bf 3}, {\bf 3})$ multiplet of massless scalars. 
The $SU(2)_{L}\times SU(2)_{R}$   symmetry is exhibited by the world-sheet currents 
\begin{equation} 
J^{\pm}(z) = :e^{\pm i\frac{2}{\sqrt{\alpha'}} X^{25}(z)}: \, ,  \qquad J^{3}(z) = \frac{i}{\sqrt{\alpha'}}\partial X^{25}(z) 
\end{equation}
and their antiholomorphic counterparts.

For open strings there are no $SU(2)$ gauge bosons in the spectrum but there is an $SU(2)$ triplet of massless 
scalars. This means that we have only global $SU(2)$ symmetry in space-time. 
  Before we  write down the corresponding currents let us introduce some notations and fix the conventions. 
The  open string mode expansion for the fields    $X^{\mu}(\sigma , \tau)$  satisfying the Neumann-Neumann boundary conditions 
reads 
$$
X^{\mu} = x^{\mu} + p^{\mu} \tau + i \sum_{n\ne 0}\frac{1}{n}\alpha^{\mu}_{n}e^{-in\tau} cos(n\sigma)  
$$
where $0\le \sigma \le \pi$ is a spatial coordinate along the string  and $\tau$ is Minkowski world sheet time variable. 
The Virasoro generators are 
$$
L_{m} =  \frac{1}{2}\sum_{n=-\infty}^{+\infty} :\alpha_{m-n}^{\mu}\alpha_{\mu n}: 
$$ 
where $\alpha^{\mu}_{0} = p^{\mu}$. 
Here and everywhere below we set $\alpha' = 1/2$. 
By mapping this theory on the upper half plane with a complex coordinate $z =  exp(  \tau  + i\sigma )$  we can further consider  
an equivalent  chiral theory  
 obtained  by the standard doubling trick. Namely, one defines $T_{zz}$ in the lower half of $z$-plane as the value of $T_{\bar z\bar z}$ 
at its image in the upper half plane under the reflection given by complex conjugation. This way we get $26$ chiral 
holomorphic fields $X^{\mu}(z)$ (by abuse of notation we will denote these fields by the same symbol $X^{\mu}$) with a mode expansion 
$$ 
\partial X^{\mu}(z) = -i \sum_{m = -\infty}^{+\infty} \frac{\alpha_{m}^{\mu}}{z^{m+1}}
$$  
and the energy-momentum  tensor of the standard form
$$
T(z) = -\frac{1}{2}:\partial X^{\mu}(z)\partial X_{\mu}(z): \, .
$$

 The Hilbert space of 
the open string compactified on the circle of self-dual radius in the $X^{25}$-direction 
  furnishes a  level one representation of an $SU(2)$ current  algebra generated by 
\begin{equation} \label{su2} 
J^{\pm}(z) = :e^{\pm i\sqrt{2}X^{25}(z)}: \, ,  \qquad J^{3}(z) = \frac{i}{\sqrt{2}}\partial X^{25}(z) \, .
\end{equation}
We use the normal ordering convention in which the zero modes are also ordered: $p^{\mu}$ stands to the right of $x^{\mu}$.
 In general it is known that the only level one ground state  representations (integrable in mathematics terminology)  of  
$SU(2)$ current algebra are those with spin $j=0$ or $j=1/2$. They both can be realized in terms of vertex operators 
(\ref{su2}) acting in the Fock space of $\alpha^{25}_{n}$ modes.  
The highest weight state for the singlet representation is the string $SL(2, {\bf R})$ vacuum $|0\rangle$, 
whereas the $j=1/2$ representation is built on the doublet of states 
$ \Bigl| \pm \frac{1}{\sqrt{2}} \Bigr> = e^{\pm \frac{i}{\sqrt{2}}x^{25}}|0\rangle$.
The representation spaces are spanned by the states 
$$
J_{-n_{1}}^{a_{1}}\cdot \dots \cdot J_{-n_{k}}^{a_{k}}|w\rangle
$$ 
where $J_{n}^{a}$ stand for the Laurent modes of $J^{a}(z)$ and 
$|w\rangle$ stand for the vacuum vectors $|0\rangle$ or $ \Bigl| \pm \frac{1}{\sqrt{2}} \Bigr>$. 
We will denote the corresponding representation spaces ${\cal F}_{0}$ and ${\cal F}_{1/2}$ respectively. 
Then the  Hilbert space of the compactified string states that we have at hand coincides with ${\cal F}_{0}$.

Let us put   $U(2)$ Chan-Paton factors on one  end of our string (say at $\sigma = 0$). 
Then we obtain a total global symmetry group isomorphic to $SU(2)\times U(2)$. 
 The Chan-Paton degrees of freedom can be coupled to the  local $SU(2)$ currents at $\sigma = 0$. 
Namely we can consider   a background  characterized by the following (total) stress energy tensor  
\begin{equation} \label{T}
T(z) = -\frac{1}{2}:\partial X_{\mu}(z) \partial X^{\mu}(z): + g \sum_{a} \sum_{n \in {\bf Z}} S^{a}J^{a}_{n} 
\end{equation} 
where $S^{a}$ are  matrices of the fundamental representation of $SU(2)$  acting in a two-dimensional complex vector 
space $E$ of  the  Chan-Paton degrees of 
freedom;  $g$ is a coupling constant. 
This background  breaks the symmetry group down to $SU(2)_{diag}\times U(1)$.
Note that the conformal dimension of the perturbation  term in (\ref{T})  is 1 so formally we are dealing with a marginal perturbation. 
However it turns out not to be truly marginal. One can show  that the coupling constant $g$ starts 
to run at higher orders of perturbation theory. 
This is in contrast with the model considered in papers \cite{CalMald} , \cite{Polchinski} 
in which one couples only the $U(1)$ subgroup  generated by the current $J^{1}(z)$ to an  end of the string. As it was shown 
in \cite{CalMald} , \cite{Polchinski} the perturbed theory is conformal for any value of the coupling constant. 

The Hamiltonian that governs the dynamics of $X^{25}$ can be rewritten solely in terms of the  $SU(2)$ currents as 
\begin{equation} 
H = \frac{1}{3}\sum_{n, a} :J^{a}_{n}J^{a}_{-n}: + g \sum_{n, a} S^{a}J^{a}_{n} \, .
\end{equation}  
At the special point $g=2/3$ we can rewrite the Hamiltonian as 
\begin{equation} \label{Sug}
H = \frac{1}{3} \sum_{n} :(J^{a}_{n} + S^{a})(J^{a}_{-n} + S^{a}): 
\end{equation}
where we dropped  an infinite additive constant.  We see that at this point the conformal symmetry is restored 
and is given by the current algebra with generators $T_{n}^{a}  = J^{a}_{n} + S^{a}$. Note that the operators 
$T_{0}^{a}$ generate the global  (on the world sheet) part of the surviving symmetry group $SU(2)_{diag}$.
The hypothesis is that the renormalization group flow brings the theory to this conformal  point. 
Assuming this we can further investigate what happens to the physical state space of the theory.


Let us show how to  construct a highest weight state of the current algebra generated by $T_{n}^{a}$'s within the original   
Hilbert space  ${\cal F}_{0}\otimes E$  that is  the string oscillator space tensored with the representation space of Chan-Paton degrees of freedom.
 The last one is  spanned  by two fermions 
denoted $\chi_{+}$ and $\chi_{-}$. We will assume that $S^{3}$ is diagonalized in this basis. 
The  $T_{n}^{a}$ highest weight state can be written as 
 
\begin{equation} \label{hw}
 | W\rangle = (:e^{\frac{i}{\sqrt{2}}X(1)}: \otimes \chi_{-} - 
:e^{-\frac{i}{\sqrt{2}}X(1)}: \otimes \chi_{+}) |w\rangle 
\end{equation}
where the operator in the brackets acts on a highest weight state $ |w\rangle $ of the current algebra generated by $J_{n}^{a}$'s. 
When we omit the vector index of $X^{\mu}$ we refer to the compactified direction $X^{25}$. 
It is convenient to introduce a notation 
$$
{\cal S} = (:e^{\frac{i}{\sqrt{2}}X(1)}: \otimes \chi_{-} - 
:e^{-\frac{i}{\sqrt{2}}X(1)}: \otimes \chi_{+}) \, .
$$
We will refer to this  expression as a singlet (or screening) operator.  
 For the term ``operator'' to make sense one should consider  ${\cal S}$ to be acting 
from the Fock space ${\cal F}$ of the modes $\alpha^{25}_{n}$  to ${\cal F}\otimes E$. 
It is not hard to check that the operator $\cal S$ satisfies the following important relation 
\begin{equation} \label{relation}
T_{n}^{a} {\cal S}  = {\cal S} J_{n}^{a} \, .  
\end{equation} 
This relation makes it obvious that the construction (\ref{hw})
 indeed gives a  $T_{n}^{a}$ current algebra highest weight state . 
The question now is what state $ |w\rangle $ should we take if the original Hilbert space ${\cal F}_{0}\otimes E$ was built on the string 
vacuum $|0\rangle$? Note that the original Hilbert space contains states whose  momenta are quantized as 
$p = n\sqrt{2}$, $n\in {\bf Z}$  
that corresponds  to integer values of the  isospin projection. Whereas the operator in (\ref{hw}) contains factors of 
$e^{\pm \frac{1}{\sqrt{2}} x}$ shifting the momenta by $\pm 1/\sqrt{2}$ and isospin projections by $1/2$. 
Assuming that the renormalization group flow cannot change the moding of momenta the only possibility we arrive at is taking 
for $|w\rangle$ the $j=1/2$ highest weight state. Then the  momenta of the representation space built on $|W\rangle$ 
is  quantized as before, although the $T^{a}_{n}$ algebra isospin is quantized in half integers. 
Strictly speaking as the states (\ref{hw}) have an infinite norm we cannot argue that we stay exactly in 
the old Hilbert space. 
So the above assumption is more a conjecture than  a rigorous result. It is of the same nature  as 
the assumptions on the RG flow in the Kondo model, which seem to be thoroughly tested (\cite{Affleck} and references therein).

The new physical state space of the string is thus built on the doublet of states 
\begin{equation} \label{pm}
 | \pm \rangle = {\cal S} \Bigl| \pm \frac{1}{\sqrt{2}} \Bigr> \, .
\end{equation}
 With this notation in mind the above assumption is that the RG flow acts inside ${\cal F}\otimes E$ and 
maps   the two-dimensional subspace $|0\rangle\otimes E$ to the subspace spanned by $|\pm\rangle$.    

Now we can calculate the new spectrum. The energies are lifted by the eigenvalue of $L_{0}$ evaluated on 
the new vacuum states $| \pm \rangle$. Since the Hamiltonian (\ref{Sug}) is given by the standard Sugawara construction 
the energy shift is determined  by the value of the Casimir operator for the $SU(2)$ fundamental representation. 
The  shift in the masses squared comes out to be equal to $1/2$ versus the tachyon mass squared being 
$-2$. (Below we will see that for some of the  more general compactifications  the tachyon mass can get  lifted all the
  way up to the zero value.) 
Thus the lowest energy state is tachyonic with the mass squared $-3/2$ and 
forms a doublet under $SU(2)_{diag}$. Below we call the corresponding particle 
a quark. There are no massless states, the triplet of scalar states and the photon acquire mass squared of the value $1/2$. 
Moreover each of those states 
splits into an $SU(2)_{diag}$ doublet because of the vacuum splitting.


So far we have discussed the situation when only one end of the string is coupled to the background. 
It is not hard to modify the whole picture above to the case when both ends are coupled. We assume that the string 
is oriented and one end carries Chan-Paton degrees in the fundamental representation and another one in the antifundamental. 
The Hamiltonian now takes the form
\begin{equation}
H = \frac{1}{3}\sum_{n, a} :J^{a}_{n}J^{a}_{-n}: + g \sum_{n, a} S^{a}J^{a}_{n} + 
\tilde g \sum_{n, a} (-1)^{n} \tilde S^{a}J^{a}_{n} 
\end{equation}
where $\tilde S^{a}$ represent the $SU(2)$ algebra action on a two-dimensional space $\tilde E$ of the 
Chan-Paton degrees at the second end ($\sigma = \pi$). This basis in $\tilde E$ will be denoted 
by $\tilde \chi_{\pm}$.   
By locality argument we assume that the theory flows to the conformal point $g=\tilde g = 2/3$ at which 
the energy-momentum tensor is given by the Sugawara construction corresponding to the currents 
$T_{n}^{a} = J_{n}^{a} + S^{a} + (-1)^{n}\tilde S^{a}$.

The construction (\ref{hw}) of  highest weight states  generalizes in a straightforward manner 
\begin{equation}
|W\rangle = {\cal S}\tilde {\cal S}|w\rangle 
\end{equation}
where 
$$
\tilde {\cal S} = (:e^{\frac{i}{\sqrt{2}}X(-1)}: \otimes \tilde \chi_{-} - 
:e^{-\frac{i}{\sqrt{2}}X(-1)}: \otimes \tilde \chi_{+}) \, .
$$
We see that now the expression ${\cal S}\tilde {\cal S} $ that acts on the ``old'' highest weight state  $|w\rangle$ contains the 
allowed momenta $p = n\sqrt{2}$ and thus one should choose $|w\rangle = |0\rangle$. As far as the whole representation space 
of the current algebra $T_{n}^{a}$ is concerned we believe that is the correct assumption. 
However, since the ``screening'' of quarks by the $X^{25}$-modes takes place locally at each end,    
the old vacuum subspace  
state $|0\rangle\otimes E\otimes \tilde E$ should flow to a subspace that is  a  tensor product in some sense of 
the vacuum subspaces at each end of the string. The last one splits into a sum of a singlet  and a triplet representation. 
More precisely the four-dimensional subspace $|0\rangle\otimes E\otimes \tilde E$ should flow to a  subspace 
spanned by the highest weight state $|s\rangle = {\cal S}\tilde {\cal S}|0\rangle$ and descendant states 
$$
T_{-1}^{\pm}|s\rangle = {\cal S}\tilde {\cal S}e^{\pm i\sqrt{2}x}|0 \rangle \, , \quad  
T^{3}_{-1} |s\rangle =  {\cal S}\tilde {\cal S}\frac{1}{\sqrt{2}}\alpha^{25}_{-1}|0\rangle \, .   
$$ 
 The above three states form a zero mass triplet representation of the global  $SU(2)_{diag}$ symmetry group.
The last one is generated by operators $T_{0}^{a}$.  

We conclude that in the case when both boundaries are coupled to the background the spectrum of the theory at the fixed 
point stays the same as that of the original free theory. This agrees with the results of \cite{CalMald}, \cite{Polchinski}.

Let us discuss now a possible space-time interpretation of the resulting model. First note that since we switched on a boundary perturbation 
the bulk central charge remains the same, $c=1$,  in the course of the RG flow  (otherwise we would be in trouble with string theory applications). 
Moreover, the resulting CFT should be describable as a free theory supplied with  new conformally invariant boundary conditions. As 
we have seen the sole effect of the flow in a sector with only one boundary coupled to the background is in the shift 
of zero mode of momentum. This can be accounted for by switching on a $U(1)$ Wilson line with the value $\theta = \pi$. 
Then on the T-dual circle we have a system of two D24 branes, call them ``I'' and ``II'',  that seat at  opposite points on the circle. 
The I-II sector of the DD boundary conditions then corresponds to the ``one boundary sector'' of our theory with the vacuum space 
spanned by $|\pm\rangle$ states (\ref{pm}). The degeneracy of the new vacuum  corresponds to the two homotopically nonequivalent 
paths of minimal lengths between the positions of the two D24-branes (two semi-circles). The unshifted I-I and II-II sectors 
can be matched with the ``two boundary'' sector having the vacuum ${\cal S}\tilde {\cal S}|0\rangle$ and with the original 
 Fock space of the compactified states built on $|0\rangle$.  The last one 
 can be thought of as a sector carrying trivial $U(1)$ Chan-Paton factors on both ends. 
(As we will see in the next section, once we introduce a one boundary sector in consideration both the two boundary and the 
trivial Chan-Paton's sectors need to be added not to violate the unitarity.)

\section{Vertex operators and 4-point functions} 
In this section we will discuss the vertex operators for the two cases considered above, i.e. 
 when only one  boundary carries  Chan-Paton factors and when both ends carry them. 
We begin with a  construction of  the $SU(2)$ current algebra  vertex operators creating the states 
$| \pm \rangle$ in the ``one boundary'' sector. This vertex operators can be written as  
\begin{eqnarray} \label{V}
 V_{\pm}(z) &=& (1-z)^{\pm 1/2} :exp \left( \frac{i}{\sqrt{2}}(\pm X(z) + X(1) ) \right): \chi_{-} - \nonumber \\
&& (1-z)^{\mp 1/2} :exp\left( \frac{i}{\sqrt{2}}(\pm X(z) - X(1) ) \right): \chi_{+} \, .	
\end{eqnarray}
This construction becomes almost obvious when rewritten with a help of the singlet operator  ${\cal S}$
$$
V_{\pm}(z) =  {\cal S} :e^{\pm \frac{i}{\sqrt{2}}X(z)}: \, .
$$

The vertex operator (\ref{V}) can be thought of as an operator creating a quark in a doublet representation out of an 
isospin zero state. 
The opposite process is governed by the vertex operator 
\begin{eqnarray} \label{V'}
 V^{\dagger}_{\pm}(z) &=& (z-1)^{\pm 1/2} :exp \left( -\frac{i}{\sqrt{2}}(\pm X(z) + X(1) ) \right): \chi_{-}^{\dagger} - \nonumber \\
&& (z-1)^{\mp 1/2} :exp\left( -\frac{i}{\sqrt{2}}(\pm X(z) - X(1) ) \right): \chi_{+}^{\dagger} \, .	
\end{eqnarray}
It can be also written as 
$$
V^{\dagger}_{\pm}(z) = :e^{\mp \frac{i}{\sqrt{2}}X(z)}:{\cal S}^{\dagger}
$$
where 
$$
{\cal S}^{\dagger} =  (:e^{-\frac{i}{\sqrt{2}}X(1)}: \otimes \chi_{-}^{\dagger} - 
:e^{\frac{i}{\sqrt{2}}X(1)}: \otimes \chi_{+}^{\dagger}) \, .
$$
We will refer to the vertex operator (\ref{V}) as a quark vertex operator and to (\ref{V'}) as an antiquark one. 

We can calculate a  four-quark tree level amplitude with these vertices. 
Up to cyclic permutations there is a single tree level  process. 
It can be represented by the following diagram 
 
\begin{figure}[h]
\begin{center}
\epsffile{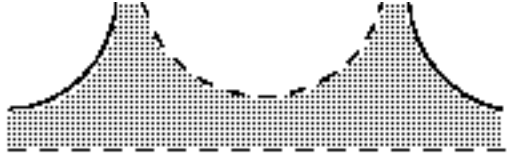}
\end{center}
\caption{  Four quark scattering}
\end{figure}
where the solid line denotes the quark boundary and the dashed line is for a nonquark boundary (singlet state).  
The  correlation function corresponding to this process is 
\begin{equation} \label{4pt}
\langle -\xi_{1} | V_{\xi_{2}}(z_{1}) V_{-\xi_{3}}^{\dagger}(z_{2}) |\xi_{4}\rangle
\end{equation} 
where the indices $\xi_{i}$  are each   $+$ or $-$. The minus signs in front of $\xi_{i}$ in (\ref{4pt}) 
are inserted so that the isospins of the incoming quarks (that are opposite of those for the antiquarks) 
are precisely the values of $\xi_{i}$. The internal momenta carried by the quarks are $p_{i} = \xi_{i} (1/\sqrt{2})$. 
 The vertex operators can be expressed as  $V_{\xi_{i}}(z) = {\cal S} v_{p_{i}}(z)$, 
$V^{\dagger}_{\xi_{i}}(z) = v_{p_{i}}(z){\cal S}^{\dagger}$ where $v_{p_{j}}(z)= :e^{ i p_{j}X(z)}:$. 
 Using these expressions and (\ref{pm}) we can rewrite (\ref{4pt}) as 
\begin{equation} \label{4pt'}
\Bigl< p_{1} \Bigr| ({\cal S}^{\dagger} {\cal S}) v_{p_{2}}(z_{1})v_{p_{3}}(z_{2}) 
({\cal S}^{\dagger} {\cal S}) 
\Bigl| p_{4} \Bigr> \, . 
\end{equation}
This expression is divergent and the source of divergence is the infinite norm of  states $|\pm\rangle$ coming from 
contractions of $X(1)$ with itself. To make sense of expression  (\ref{4pt'}) we will employ a mode cutting regularization 
in which the  operators of the form $:e^{ikX(z)}:$ are regularized as 
\begin{equation} \label{mode_cut}
:e^{ikX(z)}:_{N} = e^{k\sum_{1}^{N}\frac{\alpha_{-n}}{n}z^{n}} e^{-k\sum_{1}^{N}\frac{\alpha_{-n}}{n}z^{-n}}e^{ikx}z^{kp} \, .
\end{equation}
The product of regularized operators ${\cal S}^{\dagger}_{N}$ and  ${\cal S}_{N}$ takes the form
$$
  {\cal S}^{\dagger}_{N} {\cal S}_{N} = C_{N}( \chi_{-}^{\dagger}\chi_{-} +  \chi^{\dagger}_{+}\chi_{+}) - 
 C_{N}^{-1}( :e^{-\sqrt{2}X(1)}:_{N}\chi_{-}^{\dagger}\chi_{+} - :e^{\sqrt{2}X(1)}:_{N}\chi^{\dagger}_{+}\chi_{-})
$$
where 
$$
C_{N} = e^{\frac{1}{2} \sum_{1}^{N}\frac{1}{n}}
$$
is a constant that tends to infinity as $N\to \infty$. 
After contracting  the Chan-Paton factors  in the above expression  we end up with the formula
\begin{equation} \label{SSreg}
{\cal S}^{\dagger}_{N} {\cal S}_{N} = 2C_{N} \, . 
\end{equation}
That is the regularized product of these two operators is proportional to the identity operator.  
Using this result in the regularized expression (\ref{4pt'}),   rescaling  
 the whole correlator by $1/(4C^{2}_{N})$ (that effectively normalizes the states $|\pm\rangle$)
we obtain after   sending $N$ to infinity the following simple result 
$$
\Bigl< p_{1} \Bigr| v_{p_{2}}(z_{1})v_{p_{3}}(z_{2}) \Bigl| p_{4} \Bigr>
$$
that is just a correlation function of ``old'' vertex operators calculated for the states of nonzero momenta 
$\pm  \frac{1}{\sqrt{2}}$. This result illustrates  best that the sole effect of the RG flow is
 the shift in the vacuum momenta, i.e. the shift of the zero modes of $\partial X^{25}(z)$.

The tree level four-quark amplitude $A(\xi_{i}, k_{i})$ depends on the isospin projections $\xi_{i}$ of 
the incoming quarks and on the energy-momentum vectors  $k_{i}^{\mu}$, $\mu = 0, \dots , 24$ in the transverse
$1 + 24$ directions. Depending on the values of $\xi_{i}$ we get three possible types of amplitudes 
$A(\xi_{i}, k_{i}) = \delta^{25}(\sum_{i} k^{i}) A_{j}(s, t)$ : 
\begin{eqnarray} 
&& A_{0} (s, t) = \int_{0}^{1} \, dx (1-x)^{-s/2 - 2}x^{-t/2 - 2} \, ,  
\quad \xi_{1} = -\xi_{2} = \xi_{3} = -\xi_{4} \, \nonumber \\ 
&& A_{1} (s, t) =  \int_{0}^{1} \, dx (1-x)^{-s/2 - 1}x^{-t/2 - 2} \, ,  
\quad \xi_{1} = \xi_{2} = -\xi_{3} = -\xi_{4} \, \nonumber \\   
&& A_{2} (s, t) = \int_{0}^{1} \, dx (1-x)^{-s/2 - 2}x^{-t/2 - 1} \, ,  
\quad \xi_{1} = -\xi_{2} = -\xi_{3} = \xi_{4} 
\end{eqnarray}
where $s = -(k_{3} + k_{4})^{2}$, $t=-(k_{1} + k_{3})^{2}$. 
Evidently these amplitudes satisfy the duality relations $A_{0}(s, t) = A_{0}(t, s)$, 
$A_{1}(s, t) = A_{2}(t, s)$. The total tree level amplitude is obtained by summing up over all 
noncyclic permutations over the external states $(k_{i}, \xi_{i})$.

It is clear from the picture that different  factorizations of the process represented on  Figure 1 will 
contain as intermediate states strings with quarks running on two boundaries as well as strings with no quarks 
on either end. Thus,  there should be   vertex operators emitting 
 ``two boundary strings'' (strings carrying Chan-Paton factors on both ends)   from  the ``one boundary'' 
strings.  We should think of three  different sectors of the same theory rather than of the three  isolated cases.     
By three sectors we mean ``one boundary strings'', ``two boundary strings'', and strings with no Chan-Paton degrees 
of freedom (except for trivial $U(1)$ factors) on either end. The lowest state of the first sector is a quark whose mass 
is tachyonic, and the rest of the states are all massive $SU(2)_{diag}$ doublets. The other two sectors have identical spectra, 
containing the usual tachyon, triplet of massless scalars (a meson)  and a photon.

We start by  constructing    meson (quark-quark) vertex operators. These operators should create strings with quarks 
running on both boundaries. The corresponding diagram of a four-meson scattering looks like that depicted on 
Figure 1 with all solid line boundaries. 
It follows from the results of the previous section  that a general physical state 
in the case when both ends of the string carry Chan-Paton degrees of freedom has a form 
${\cal S}\tilde {\cal  S} |s\rangle $ where    $|s\rangle$ is a physical state in the ``old'' Fock space. 
Thus, we see that the vertex operators we are looking for should satisfy 
\begin{equation} \label{SSvop}
\lim_{z\to 0} V^{a}(z) {\cal S}\tilde {\cal  S} |0\rangle = \lim_{z\to 0} { \cal S}\tilde {\cal  S} J^{a}(z) |0\rangle \, . 
\end{equation}   
They create a massless particle  in the adjoint representation of $SU(2)$ 
 to which we will refer as a meson. 
If we are looking for a vertex operator that emits a meson from the boundary $\sigma = 0$ we can 
omit the operator $\tilde S$ from the above formulas. Looking at formula (\ref{relation}) we see that the modes 
$T_{n}^{a}$ satisfy the necessary relation. However the local operator with these modes is singular, it contains a 
delta function. To obtain  a regularized version of  $V^{a}(z)$ we may start with the following formal expression 
\begin{equation} \label{mesonV}
V^{a}(z) = {\cal S} J^{a}(z) {\cal S}^{\dagger} \, . 
\end{equation}
One way of obtaining  this expression is by fusing the vertices $V_{\pm}(z)$, $V^{\dagger}_{\pm}(z)$ we 
obtained before.  To be careful one has to  fuse the regularized vertices and keep all terms in the regularized 
expression after the fusion. 
Another, much more practical  way to  make sense of  the expression (\ref{mesonV}) is by defining its action on a general physical state 
${\cal S}|s\rangle$. In the expression ${\cal S} J^{a}(z) ({\cal S}^{\dagger}{\cal S}) |s\rangle$ we may start by 
regularizing $({\cal S}^{\dagger}{\cal S})$ as in  (\ref{mode_cut}) and then use (\ref{SSreg}). We see that after a proper rescaling 
 the regulated vertex satisfies \ref{SSvop}.

In order to calculate an $n$-point meson scattering amplitude we regularize the correlator by cutting the modes as 
\begin{eqnarray*} 
&&\langle 0| {\cal S}^{\dagger} ({\cal S}J^{a_{1}}(z_{1}){\cal S}^{\dagger})\cdot \dots \cdot 
({\cal S}J^{a_{n}}(z_{n}){\cal S}^{\dagger}) {\cal S}|0\rangle_{reg} = \nonumber \\
&& \langle 0| ({\cal S}^{\dagger}_{N} {\cal S}_{N})J^{a_{1}}(z_{1})_{N} ({\cal S}^{\dagger}_{N} {\cal S}_{N})\cdot \dots \cdot 
J^{a_{n}}(z_{n})_{N}({\cal S}^{\dagger}_{N} {\cal S}_{N})|0\rangle \, .
\end{eqnarray*}
Now formula (\ref{SSreg}) can be applied to each factor $({\cal S}^{\dagger}_{N} {\cal S}_{N})$ that results in an 
overall divergent numerical factor. 
After rescaling and taking the limit  $N\to \infty$ we 
obtain that the contribution of the  compactified modes  boils down to  a correlator of currents 
$$
\langle 0| J^{a_{1}}(z_{1})\cdot \dots \cdot J^{a_{n}}(z_{n})|0\rangle  
$$
that can be easily evaluated by algebraic methods.  
Note that in the above discussion for brevity we omitted the operators $\tilde {\cal S}$ present in the ``two boundary'' string  states. 
Contractions between ${\cal S}$ and $\tilde {\cal S}$ lead to trivial constant factors.  The derivation above 
can be modified to include  operators  $\tilde {\cal S}$ in a straightforward way. The result is the same. 
All other amplitudes can be calculated in the same fashion. For example two quarks - two mesons amplitude 
contains a correlator 
$$
\langle p_{\infty}|{\cal S}^{\dagger} ({\cal S}J^{a_{1}}(z_{1}){\cal S}^{\dagger}) ({\cal S}J^{a_{2}}(z_{2}){\cal S}^{\dagger})
{\cal S} |p_{0}\rangle
$$
where $p_{0}$ and $p_{\infty}$ are $\pm 1/{\sqrt{2}}$ according to quark isospin projections. 
After regularization and rescaling we have 
$$
\langle p_{\infty}|J^{a_{1}}(z_{1})J^{a_{2}}(z_{2})| p_{0}\rangle \, .
$$
We see that whenever we have states in the ``one boundary'' sector we should include a zero momentum shift for the compactified mode. 
With the above examples in mind constructions of all other possible amplitudes should be quite clear.

\section{Compactification of two dimensions. Relation with Kondo model}

Next we would like to discuss the compactification of two dimensions on circles  each having the self-dual radius. 
Obviously in this case we obtain an  $SU(2)\times SU(2)$ current algebra. Take the representation $S^{a}$ to 
be a $(1/2, 1/2)$-representation.  
It is clear  then how to extend our considerations for a single $SU(2)$ to the case at hand. For example 
the analog of the construction (\ref{hw}) now reads  
\begin{equation} \label{so4hw}
| W\rangle = (:e^{\frac{i}{\sqrt{2}}X^{25}(1)}: \otimes \chi_{-} - 
:e^{-\frac{i}{\sqrt{2}}X^{25}(1)}: \otimes \chi_{+})( :e^{\frac{i}{\sqrt{2}}X^{24}(1)}: \otimes \chi_{-}' - 
:e^{-\frac{i}{\sqrt{2}}X^{24}(1)}: \otimes \chi_{+}' ) |w\rangle 
\end{equation}
where the Chan-Paton space is assumed to be spanned by vectors $\chi_{\pm}\otimes \chi'_{\pm}$. (Do not confuse 
the $\chi'$ states with the states $\tilde \chi$ used above for the two boundaries case.) 
By the same reasoning as before one should take $|w\rangle = \Bigl| \pm \frac{1}{\sqrt{2}}, \pm \frac{1}{\sqrt{2}} \Bigr>$ 
(now the state is marked by the values of momenta  $p^{25}$ and $p^{24}$ respectively). The ground energy shift amounts to 
shifting the tachyon mass squared by 1 unit up.

It is  instructive to recast the construction of a new highest weight state  in terms of free fermions.
The $SU(2)\times SU(2)$ current algebra  is isomorphic to the $SO(4)$ current algebra and the whole picture 
above can be represented in terms of 4 chiral Neveu-Schwarz (NS) fermions $\psi^{i}(z)$. 
 The Chan-Paton factors $\chi_{i}$, $i=1,\dots , 4$ transform in a vector representation.  
The $SO(4)$ currents have the form
\begin{equation} \label{so4}
J_{ij}(z) = -i :\psi_{i}(z) \psi_{j}(z):  
\end{equation}
The mode expansion for $\psi^{i}(z)$ reads
$$
\psi^{i}(z) = \sum_{r\in {\bf Z} + 1/2} b_{r}^{i} z^{r}  
$$
and the Fock vacuum is defined as 
$ b_{r}^{i} |0\rangle = 0\, , \, r>0 $. The whole Fock space splits into a direct sum of two irreducible 
representations of the current algebra (\ref{so4}). The subspace of even fermion number corresponds 
to the singlet representation and the  odd fermion number subspace is built on a vacuum space 
spanned by the vectors $b_{-1/2}^{i}|0\rangle$ transforming in a vector representation of $SO(4)$. 
Notice that in (21) we have chosen half integer moded NS fermions rather than integer moded Ramond
fermions. The reason for this is that the even fermion sector of NS fermions corresponds to integer
moding of the bosonic $X^{24}$ and $X^{25}$ momenta, which is required by the self dual radius
compactification. In contrast, R fermions correspond to half integer moding of the bosonic
momenta.
 The highest weight state construction  (\ref{so4hw}) in the fermionic language reads as 
$$
|W\rangle = ( \sum_{i = 1}^{4} \psi^{i}(1)\chi^{i} ) |w\rangle 
$$ 
and staying in the same representation space requires the choice $|w \rangle = b^{i}_{-1/2}|0\rangle$.


Let us consider now a more interesting situation that is directly related to the Kondo model. 
Consider again the compactification on two circles of self-dual radii. But instead of 
coupling the whole $SO(4)\approx SU(2)_{L}\times SU(2)_{R}$ to the Chan-Paton factors let us couple now only one 
of the $SU(2)$ subalgebras. (The subscripts $L$ and $R$ is just a convention here distinguishing two copies of 
$SU(2)$'s.)  
The answer for the resulting flow is of course the same as the one we 
considered before for the one direction compactified. However the whole picture can be fermionized now.   
Let us work now with two chiral Weyl fermions $\psi^{\alpha}(z)$, $\alpha = 1,2$. 
The currents of  $SU(2)_{L}$ read 
$$
J^{a}(z) = \frac{1}{2}:\psi^{\dagger\alpha }(z)\sigma^{a}_{\alpha \beta} \psi^{\beta}(z): 
$$
where $\sigma^{a}$ are Pauli matrices. Currents of $SU(2)_{R}$ chosen in a Cartan basis are 
\begin{eqnarray*}
&& K^{3}(z)  = \frac{1}{2}\sum_{\alpha = 1,2}:\psi^{\dagger}_{ \alpha}(z) \psi^{\alpha}(z) : \, , \nonumber \\
&& K^{+}(z) = :\psi^{\dagger }_{1}(z)\psi^{\dagger}_{ 2}(z): \, , \qquad K^{-}(z) =:\psi^{2}(z)\psi^{1}(z): \, .  
\end{eqnarray*}
The operator $K^{3}_{0}$, where zero refers to the Laurent mode, is just one half of the charge operator. 
We thus start with a perturbed stress energy tensor 
$$
T(z) = \frac{1}{2}:\psi^{\dagger}_{\alpha}(z)\partial \psi^{\alpha} (z): + g \sum_{a}J^{a}(1)S^{a} 
$$
This formula  gives  precisely  the Kondo model stress-energy in the holomorphic representation.
In terms of the currents we have 
$$
T(z) = \frac{1}{3}:J^{a}(z)J^{a}(z): + :K^{3}(z)K^{3}(z): + g \sum_{a}J^{a}(1)S^{a}
$$
i.e. we have the Sugawara construction for the group $U(2)$ and the old perturbation term that involves only 
the $SU(2)$ subgroup.

We can work out now the new highest weight state in terms of fermion modes and the $\chi^{\alpha}$ states. 
The mode expansions for $\psi$ and $\psi^{\dagger}$ are of the form 
\begin{eqnarray}
&&\psi^{\alpha} = \sum_{r\in {\bf Z} + 1/2} b_{r}^{\alpha} z^{ r} \, , \nonumber \\
&&\psi^{\dagger\alpha} = \sum_{r\in {\bf Z} + 1/2} b^{\dagger \alpha} z^{- r}
\end{eqnarray}
and commutation relations are
$$
\{ b_{r}^{\alpha} , b^{\dagger \beta}_{s} \} = \delta_{r, s}\delta^{\alpha  \beta} \, .
$$ 
The standard  vacuum state $|0\rangle$ in the Fock space ${\cal F}^{ferm}$ is defined as 
$$
b_{r}^{\alpha} |0\rangle = 0\, , \, r>0 \, \qquad b_{ r}^{\dagger \alpha} |0\rangle = 0\, , \, r<0 \, .
$$
We can construct  now two $SU(2)_{L}$ singlet operators 
\begin{eqnarray}
&& {\cal S} = \psi^{1}(1)\otimes \chi_{+} + \psi^{2}(1)\otimes \chi_{-} \, , \nonumber \\
&&  \bar {\cal S} = \psi^{\dagger  1}(1)\otimes \chi_{-} - \psi^{\dagger  2}(1)\otimes \chi_{+} \, .  
\end{eqnarray}
If the Hilbert state we begin with has states of even fermion number only (that would be the case 
if we are interested in the string theory application) then we have four possible doublets of $SU(2)_{L}$
$$
 {\cal S}b_{-1/2}^{\alpha}|0\rangle \, , \qquad {\cal S}b^{\dagger \alpha }_{  1/2}|0\rangle \, , \qquad 
 \bar {\cal S}b_{-1/2}^{\alpha}|0\rangle \, , \qquad \bar {\cal S}b^{\dagger  \alpha }_{ 1/2}|0\rangle \, .
$$
However, the second $SU(2)$ permutes these multiplets. The true vacuum subspace should be invariant under  
$SU(2)_{R}$. It is not hard to find that the correct vacuum space is a single  $SU(2)_{L}$ doublet 
spanned by 
\begin{eqnarray}
&&|-\rangle = ({\cal S}b^{\dagger 2}_{1/2} + \bar {\cal S}b^{1}_{-1/2})|0\rangle \, , \nonumber \\
&& |+ \rangle = ({\cal S}b^{\dagger 1}_{ 1/2} -  \bar {\cal S}b^{2}_{-1/2})|0\rangle \, .
\end{eqnarray}  
Hence, we see that the vacuum subspace transforms now in a chiral spinor representation of the whole $SO(4)$ group. 
Also the moding of the new isospin operator $T^{3}_{n} = J^{3}_{n} + S^{3}$ got shifted by $1/2$. This 
means that the new representation space is described in terms of free Ramond fermions. This fits perfectly 
with the general expectation that the boundary RG flow results only in a change of boundary conditions  
for the fields.

Conversely, suppose one starts with a Hilbert space with odd fermion number. 
(Although we do not see a string theory setup for this situation, it naturally arises in the Kondo model).
We are forced then to start with a four-dimensional vacuum space spanned  by $b^{\alpha}_{-1/2}|0\rangle$, 
$b^{\dagger \alpha}_{1/2}|0\rangle$. The new vacuum subspace then is invariant  under 
 $SU(2)_{L}$ and is a doublet with  respect to $SU(2)_{R}$. It is spanned by 
 ${\cal S}|0\rangle$ and  $\bar {\cal S}|0\rangle$. (As we started from a four-dimensional vacuum space 
times a two-dimensional Chan-Paton space and ended up with a two-dimensional vacuum,  presumably
the rest of the  states flow to descendants.)  Again we end up with Ramond fermions. 
Two conformal towers, that of $SU(2)_{L}$ and $SU(2)_{R}$ being originally glued as 
$$
(\mbox{integer isospin}, \mbox{even fermion number}) \oplus (\mbox{half-integer isospin}, \mbox{even fermion number})
$$ 
get reshuffled after the flow resulting in  
$$
(\mbox{half integer isospin}, \mbox{even fermion number}) \oplus 
(\mbox{integer isospin}, \mbox{odd fermion number}) \, .
$$ 
These results regarding the Kondo model are originally due to Affleck \cite{Affleck}, 
we just discuss  them here from  a somewhat different point of view.


\section{More general compactifications} 

In this section we consider $n$-dimensional compactification on   maximal tori of $SO(2n)$  groups and discuss 
other possible  setups.
We refer the reader to  paper \cite{curralg_review} for a review of the standard material about vertex operator 
constructions that we use in this section. 

We start with a discussion of $SO(2n)$ groups and Chan-Paton degrees in the vector representation. 
Let $e_{i}$, $i=1, \dots , n$ be a standard ortonormal basis in ${\bf R}^{n}$ and   
let $\Lambda_{R} \in {\bf R}^{n}$ be $SO(2n)$  root lattice generated by the  roots $\pm e_{i} \pm e_{j}$, $i\ne j$. 
(In this section we will stick to the normalization in which the simple  roots have length $\sqrt{2}$. ) 
Consider  a level one representation of the $SO(2n)$  current algebra by means of 
the vertex operator construction: 
\begin{equation} \label{so(n)}
E^{\alpha \pm}(z) = :e^{\pm \alpha_{j} \cdot X^{j}(z)} : c^{\alpha}(p) \, , \qquad H^{j}(z) = i \partial X^{j}(z) 
\end{equation}
where $j$ runs from $1$ to $n$, $\alpha \in \Lambda_{R}$ are positive roots,  
$c^{\alpha}(p)$ are Klein cocycle factors. The Laurent modes of  $E^{j\pm}(z)$ give us 
the ladder operators $E_{n}^{\alpha \pm}$ corresponding to  roots and the modes $H_{n}^{j}$ are Cartan operators of 
our current algebra. The possible momenta in the Fock space of modes $\alpha^{j}_{n}$, $j=1, \dots , n$ are constrained to 
lie in   the weight lattice $\Lambda_{W} = \Lambda_{R}^{*}$. For the groups $SO(2n)$ the weight lattice contains  four 
cosets $\Lambda_{W}/\Lambda_{R}$: 
$$
\Lambda_{W}(SO(2n)) = \Lambda_{R}\cup (\lambda_{v} + \Lambda_{R})\cup (\lambda_{s} + \Lambda_{R}) \cup 
(\lambda_{\bar s} + \Lambda_{R}) 
$$  
where $\lambda_{v}$, $\lambda_{s}$ and $\lambda_{\bar s}$ are the three minimal weights of $SO(2n)$ corresponding 
to the vector, spinor and conjugate spinor representations ($v$, $s$, $\bar s$) respectively. These cosets give rise to four possible 
highest weight representations of $SO(2n)$ current algebra built on the  highest weight states $|0\rangle$ - the Fock vacuum, 
$|\lambda\rangle = e^{i\lambda_{j}x^{j}}|0\rangle$ where $\lambda = \lambda_{v}, \lambda_{s}, \lambda_{\bar s}$.  
Proceeding as before we couple the currents (\ref{so(n)}) at one end of the string to the Chan-Paton degrees of freedom 
transforming in a vector representation of $SO(2n)$. The perturbed stress energy tensor expressed in terms of the currents 
reads 
\begin{eqnarray}
&& T(z) = \frac{1}{2(2n -1)} :\sum_{j=1}^{n} H^{j}(z)H^{j}(z) + \sum_{\alpha > 0 } (E^{\alpha}(z)E^{-\alpha}(z) + 
E^{-\alpha}(z)E^{\alpha}(z)): + \nonumber \\ 
&& g(\sum_{j}H^{j}(1)S^{j} + \sum_{\alpha > 0 } (E^{\alpha}(1)S^{-\alpha} + E^{-\alpha}(1)S^{\alpha} )   
\end{eqnarray}
where $S^{j}$, $S^{\pm \alpha}$ are matrices of the  $SO(2n)$ defining (vector) representation written in a Cartan basis. 
At the point $g= 1/(2n -1)$ one can complete the square and get a Sugawara construction with the new  $SO(2n)$ currrent algebra 
generators having modes $T_{n}^{j} = H_{n}^{j} + S^{j}$, $T_{n}^{\pm \alpha} = E^{\pm \alpha}_{n} + S^{\pm \alpha}$. 
The singlet operator has now the form 
\begin{equation} \label{singletop}
{\cal S} = \sum_{j=1}^{n} :e^{i X^{j}(1)}:\otimes \chi_{j}
\end{equation}
where $ \chi_{j}$ are Chan-Paton factors transforming in the vector representation. Also note that the operators 
$e^{i X^{j}(1)}$ have the momenta corresponding to fundamental weights of the vector representation. 
In order to stay in the same Hilbert space after the RG flow we have to ensure that the momenta entering the new highest 
weight state are all in the root lattice $\Lambda_{R}$. This condition is satisfied if one takes 
\begin{equation} \label{newhw}
|W\rangle \equiv |k\rangle  = {\cal S} e^{ix^{k}}|0\rangle 
\end{equation}
i.e. chooses $|w\rangle$ corresponding to the vector representation.  It is not hard to calculate the energy shift 
(see the general formula (\ref{shift}) below). For all $n$'s we have 
\begin{equation} \label{vect_shift}
(\Delta E)_{vector} = \frac{1}{2}
\end{equation}
which  means that the tachyon mass squared gets shifted half way up, i.e. from $-2$ to $-1$.
In the case when both boundaries carry Chan-Paton indices one would expect a complete lift of the tachyon mass. 
However, as it was already discussed in the $SU(2)$ case, the lifted states are descendants and the true vacuum 
(that will inevitably show up in other scattering  channels) is tachyonic. In fact for the ``two boundary'' 
states the spectrum is the same as that of the original theory.

The construction of the singlet operator (\ref{singletop})  and the new highest weight state (\ref{newhw}) 
generalizes straightforwardly  to the case of arbitrary simply-laced simple Lie algebra. If we set aside for a moment 
the consistency conditions coming from string theory we have the following general pattern. Let $\bf g$ be a simply laced 
simple Lie algebra. Choose a minimal fundamental weight  such that the corresponding 
highest weight representation of $\bf g$ gives rise to an integrable representation of the  current algebra. 
 Let $\lambda_{i}$ be  the fundamental weights of this representation. 
Assume also that the Chan-Paton factors transform in the conjugated  representation. We then write 
a singlet operator as 
\begin{equation} \label{genS} 
{\cal S} = \sum_{j}:e^{i\lambda_{j}\cdot X(1)}:\chi_{j} \, . 
\end{equation}
One should choose a suitable minimal representation with weights $\lambda'_{j}$ of $\bf g$ for $|w\rangle$  so that 
the sums $\lambda_{j} + \lambda_{k}'$ all belong to the root lattice. As it is well known (see for example \cite{curralg_review}) 
there is a correspondence between  $ Z(G)$ - the center of the universal covering group  whose Lie algebra is the compact real 
form of $\bf g$,  the factor group  $ \Lambda_{W}/\Lambda_{R}$ and the cominimal weights of $\bf g$ 
plus the  zero weight (a fundamental weight is called minimal if its dual Coxeter label is one, and is called 
cominimal if its Coxeter label is one; see \cite{group_th}). 
This correspondence holds  even for non simply-laced groups. 
Thus, we see that the cosets of $\lambda'_{i}$ and 
$\lambda_{i}$ should represent inverse elements  in the group $ Z(G)$. 
The construction of the  vacuum subspace in the two boundary sector should satisfy the same requirement of 
having momenta in the root lattice. Again the choice of the new minimal weight can be deduced from the structure  of 
the group $Z(G)$.

The ground energy shift can be expressed via values of Quadratic Casimir operators as  
\begin{equation} \label{shift}
\Delta E = \frac{C_{v}}{2( k + C_{a})}  
\end{equation}
where $C_{a}$ and $C_{v}$ are  the values of the quadratic Casimir for the adjoint representation and the representation in 
which the Chan-Paton degrees of freedom transform.

For the remaining cases of simply laced groups and minimal representations  
 there is of course a  usual question  of weather the Chan-Paton degrees are introduced correctly, i.e. in a way consistent 
 with  unitarity and factorization of amplitudes. It is well known (\cite{GSW}) that Chan-Paton degrees of freedom can be introduced 
consistently only for the groups $U(n)$, $SO(n)$ and $USp(2n)$ and  Chan-Paton indices in defining representation. 
Note however that the usual restrictions come from considerations 
of the two boundary sector and in the case at hand 
the Chan-Paton degrees get completely absorbed into the new vacuum, i.e. they enter only in ${\cal S}$ and $\tilde {\cal S}$ 
combinations. So we do not think that the usual factorization arguments are applicable in our situation. Thus, in principle we 
can consider more general setups and representations.

As an example consider now $SO(2n)$ groups with Chan-Paton indices in the spinor representation ${s}$. 
The center $Z(SO(2n))$ is isomorphic to the group ${\bf Z}_{4}$ when $n$ is odd and to ${\bf Z}_{2}\times {\bf Z}_{2}$ 
when $n$ is even. Let us first consider the case of even $n$'s.   
 If $\lambda_{\alpha}^{\bar s}$, $\alpha=1, \dots , 2^{n-1}$ are fundamental weights of the conjugated  
spinor representation ${\bar s}$ then we have 
$$
|W\rangle_{1 bound.} \equiv |\alpha\rangle = (\sum_{\beta} :e^{i\lambda_{\beta}^{\bar s} \cdot X(1)}:\chi_{\beta}) 
e^{i\lambda_{\alpha}^{\bar s}\cdot x}|0\rangle \, .
$$
The resulting energy shift can be calculated by formula (\ref{shift}) and is equal to 
\begin{equation} \label{spin_shift}
(\Delta E)_{spinor} = \frac{n}{8} \, . 
\end{equation}
Thus, for $SO(4)$ we get the familiar $1/4$ shift, for $SO(16)$ the tachyon mass gets lifted all the 
way up to zero and for $SO(32)$ it becomes massive. All states split into a spinor multiplets. 
So much for the one boundary sector. When we couple both boundaries we have Chan-Paton factors $\chi_{\alpha}$
transforming in the  ${s}$ representation running on one end and 
$\tilde \chi_{\alpha}$  in the  ${\bar s}$ representation 
on another end. To satisfy our usual requirement of having all momenta in the root lattice 
we have to choose the vector representation for the new highest weight state: 
$$
|W\rangle_{2 bound.} \equiv |j\rangle =  (\sum_{\beta} :e^{i\lambda_{\beta}^{\bar s}\cdot X(1)}:\chi_{\beta}) 
 (\sum_{\sigma} :e^{i\lambda_{\sigma}^{ s} \cdot X(-1)}:\tilde \chi_{\sigma})e^{i\lambda^{v}_{j}\cdot x}|0\rangle 
$$
and the energy shift in the two boundary sector is $1/2$. 
To summarize we obtain Hilbert spaces built using all of the four minimal representations: 
${s}$ and ${\bar s}$ in the one boundary sector, ${v}$ in the two boundary sector 
and the singlet representation in the sector with trivial Chan-Paton indices on both ends.

The considerations for the groups $SO(2n)$ with odd $n$ are parallel to the ones made above. 
One should only take into account that $[\lambda_{s}] + [\lambda_{\bar s}] = [0]$ in the group 
$\Lambda_{W}/\Lambda_{R}\cong {\bf Z}_{4}$ where square brackets stand for the corresponding 
cosets. Thus, we have the same shifts in the one boundary sectors (the only difference is that 
Chan-Paton factors coupled to the boundary have the same chirality as the resulting highest weight representation 
versus the opposite situation for even $n$'s). In the two boundary sector carrying Chan-Paton factors of 
opposite chirality on the two ends we have no lift in the ground state energy. 
Note that in principle one can  also consider  two boundary sectors having Chan-Paton factors of the same chirality 
on both ends. The corresponding energy shifts are always given by either formula (\ref{vect_shift}) or (\ref{spin_shift}).

Let us make here a remark about fermionization. For the case of the vector representation of $SO(n)$ there is 
an equivalent  picture of the flow in the Fock space  of $n$ real NS fermions. 
For the spinor representation although we do not know if there is  a simple picture of the flow that stays inside the 
Fock space of the vectorial NS fermions (except for the $SO(4)$ case considered above) 
we can always interpret the flow as a change from NS to R boundary conditions. 
For the group $SO(8)$ due to the famous triality property 
one can  alternatively recast everything in terms of spinorial fermions with NS boundary conditions \cite{E8}. In that case, 
if we start  with the even fermion number subspace in the corresponding Fock space, 
then we flow in that subspace  times the Chan-Paton factors and 
 the final result is equivalent to  the odd fermion number subspace in the Fock  space.

Let us now turn to the case of $SU(n)$ groups. The center in this case is isomorphic to ${\bf Z}_{n}$. 
This means that the group of cosets $\Lambda_{W}/\Lambda_{R}$ is cyclic, generated by the coset corresponding to 
the highest weight  of the fundamental representation. All of the elements of ${\bf Z}_{n}$ give rise 
to some highest weight representation of $SU(n)$ current algebra. The construction of the new highest weight states 
follows the general pattern discussed above and 
we will work out explicitly only the $SU(3)$ case. Let us label the Chan-Paton factors by the minimal representations: 
$\chi^{\bf 3}_{a}$, $\chi^{\bar {\bf 3}}_{a}$. Then one has two types of singlet operators, ${\cal S}^{\bf 3}$ and 
${\cal S}^{\bar {\bf 3}}$ resulting after fusing 
these factors with the twist operators $:e^{i\lambda_{a}X(1)}:$ of the conjugated representations. It is convenient 
to introduce  a trivial singlet operator ${\cal S}^{\bf 0}=1$  that corresponds to  the boundary carrying trivial Chan-Paton factors. 
Then, we have nine  sectors in the theory (nine   boundary conditions) the vacuum subspaces of which 
are constructed by using the combinations ${\cal S}^{A}\tilde {\cal S}^{B}$ where $A, B = {\bf 0}, {\bf 3}, \bar {\bf 3}$.
More explicitly we have 
\begin{equation} \label{states}
|k\rangle_{AB}={\cal S}^{A}\tilde {\cal S}^{B}e^{i\lambda_{k}^{C}\cdot x}|0\rangle
\end{equation}
where $\lambda_{k}^{C}$ are weights of the minimal representation  $C$ such that the corresponding cosets in 
$\Lambda_{W}/\Lambda_{R}$ satisfy $[\lambda^{A}] + [\lambda^{B}] + [\lambda^{C}] = 0$.
  As usual the tilted quantities refer to the $\sigma = \pi$ end of the string and the untilted ones to  
$\sigma =0$. 
The sectors corresponding to $A = {\bf 0}, B={\bf 0}$, $A={\bf 3}, B = \bar {\bf 3}$,  and $A=\bar {\bf 3}, B =  {\bf 3}$ 
have the old, unshifted energy spectrum, whereas other sectors give rise to triplets of vacuum 
states and   momenta shifted by $1/3$. 
 These shifts can be accounted for by switching on the appropriate 
Wilson lines on the torus ${\bf R}^{2}/\Lambda_{W}$. The  picture becomes more transparent on 
the T-dual torus ${\bf R}^{2}/\Lambda_{R}$. Let us put three distinct $D25$ branes on the points 
corresponding to the positions of the minimal (and zero) weights: 
$$
\Lambda^{\bf 0} = (0,0) \,,  \qquad \Lambda^{\bf 3} =-\Lambda^{\bar {\bf 3}} = (\frac{1}{3}, \frac{1}{3}) \, 
$$ 
written in the root basis.
 Oriented strings with DD boundary conditions stretching between these branes fall into nine distinct sectors  
that can be matched with the spectra of above representations $|k\rangle_{AB}$ 
if  we adopt a rule that strings stretching from the brane marked by 
the highest weight $\Lambda^{A}$ to the one marked by $\Lambda^{B}$ correspond to the combination of a pair of  singlet operators 
with $A' =  A$, $B' = \bar B$. This rule is quite natural if one looks at the  momenta carried by 
the states $|k\rangle_{AB}$.
The triple degeneracy of the shifted vacuum states corresponds to three homotopically inequivalent paths 
of minimal length stretching between each pair of distinct branes. This results from the peculiar positions of the 
branes that actually seat at the fixed points of  ${\bf Z}_{3}$ action, defined on the roots $r_{1}$, $r_{2}$ as  
$r_{1}\to r_{2}$, $r_{2}\to -r_{1} -r_{2}$.

\section{Discussion and future directions}
We have considered above various setups in which the open string vacuum state is probed by various 
almost marginal  boundary perturbations of the world sheet CFT. The resulting RG flow brings the theory to 
a nontrivial IR fixed point.  Theories at the fixed point all possess new instabilities  of the tachyon  
type. The situation is similar to the partial tachyon condensation through creation of lower dimensional branes 
considered in \cite{Sen1}, \cite{relevance}. 

We would like to propose here a possible brane interpretation of the RG flows we considered. 
First note that for the two boundary sector that can be considered by itself, the initial system 
in the case of compactification on $SU(N)$ maximal torus with $U(N)$ Chan-Paton factors added is just 
a system of $N$ D25 branes wrapped on the corresponding torus. After the flow the Chan-Paton factors are completely absorbed, 
 the new vacuum is nondegenerate and the spectrum is the same as for a single D25 brane. This suggests that the boundary RG flow at hand 
 describes some sort of a merging  of  $N$ D25 branes into a single D25 brane.

One can also try to interpret  the sectors with different Chan-Paton factors on two ends  in terms of systems of D-branes. 
For example in the $SU(2)$ case  before switching on the boundary perturbation we can consider a system of 
a single D25 brane and two Dp-branes, $p\ge 1$ on top of each other wrapped on the circle of self-dual radius (for $p\ne 25$ one can easily modify 
the considerations above by imposing the appropriate boundary conditions on the string coordinates transversal to the circle).
The sector with $U(1)$ Chan-Paton factors on one end and with $U(2)$ on another then corresponds to strings stretched between 
the D25 brane and the two Dp-branes. After switching on the background on the two Dp-branes the system flows to a configuration 
that in the T-dual picture can be described as  a single D24 brane and a single D(p-1)-brane sitting at the radially symmetric  
points on the circle.

In the case of  $SU(3)$ group at the end point of the flow we also arrive at a very symmetric configuration of the branes that are located 
at the fixed points  of a $T^{2}/{\bf Z}_{3}$ orbifold. The connections of the boundary flows we consider  
with orbifolds need to be better understood.

As was already pointed out in the introduction  the process of tachyon condensation  causes  the reduction of open strings degrees of freedom. 
Formally the number of boundary degrees of freedom in BCFT can be measured by boundary entropy $g$ \cite{AfflLud2}, \cite{AfflLud3}. 
 The values of $g$ for $c=1$ conformal field theories 
with Dirichlet  and Neumann  boundary conditions were calculated in \cite{Eliezer}. It would be interesting 
to calculate explicitly  and compare  the boundary entropies   for  the initial and the end points of the RG flows that 
we considered.

Finally let us mention possible connections with superstring theories. One can speculate that the bosonic 
string theory is dynamically driven by tachyon condensation process to some  supersymmetric theory 
or a fermionic theory related to a supersymmetric one.
In the case of $SO(32)$ theory the flow from NS to R fermions shows a  
 possible way  of how various sectors in the spectrum of  type I or heterotic superstring theories 
can be dynamically generated in the bosonic string. 
One can try more sophisticated schemes by using $SO(16)$ or $SO(8)$ subgroups and coupling them to 
various Chan-Paton factors.  We leave these questions for a future investigation.

\begin{center} {\bf Acknowledgments} \end{center} 
One of the authors(K.B) would like to thank the Institute for Advanced Study in Princeton
for its hospitality during the fall of 1999, where some of this work started. He would also
like to thank Dung-Hai Lee for references on the Kondo problem.
Both authors would like to acknowledge a useful discussion with E.~Rabinovici.

\end{document}